\documentclass[twocolumn,showpacs,preprintnumbers,amsmath,amssymb,prb]{revtex4-1}

\usepackage{graphicx}
\usepackage{color}

\begin{document}

\title{
Anomalous dielectric response in the dimer Mott insulator 
$\kappa$-(BEDT-TTF)$_2$Cu$_2$(CN)$_3$ 
}

\author{
Majed Abdel-Jawad
}
%\email[Email me at:]{
%jawad@riken.jp
%} 

\author{
 Ichiro Terasaki
}

\email[Email me at:]{
terra@cc.nagoya-u.ac.jp
} 
\altaffiliation[Present address:]{
Department of Physics, Nagoya University, 
Nagoya 464-8602, Japan
}

\affiliation{
Department of Applied Physics, Waseda University, 
Tokyo 169-8555, Japan
}

\author{
Takahiko Sasaki
}

\author{
Naoki Yoneyama
}

\affiliation{
Institute for Materials Research, Tohoku University, 
Sendai 980-8577, Japan,\\
also,
Japan Science and Technology Agency, CREST, Tokyo 
102-0075, Japan
}

\author{
Norio Kobayashi
}

\affiliation{
Institute for Materials Research, Tohoku University, 
Sendai 980-8577, Japan
}

\author{
Yoshiaki Uesu
}

\affiliation{
Department of Physics, Waseda University, 
Tokyo 169-8555, Japan
}

\author{
Chisa Hotta
}

\affiliation{
Department of Physics, Kyoto Sangyo University, 
Kyoto 603-8555, Japan
}

\begin{abstract}
We have measured and analyzed the dielectric constant of the dimer
Mott insulator $\kappa-$(BEDT-TTF)$_2$Cu$_2$(CN)$_3$,
which is known as a playground for a spin-liquid state.
Most unexpectedly, this particular organic salt has nontrivial
charge degrees of freedom, being characterized by 
a relaxor-like dielectric relaxation below around 60 K.
This is ascribed to the charge disproportionation within 
the dimer due to the intersite Coulomb repulsion.
A possible microscopic model is suggested and discussed.
\end{abstract}

\maketitle
\section{Introduction}
The electric dipole is a fundamental concept in 
describing the response of a material to an electric field.\cite{review1}
Spontaneously-emerging electric dipoles are of particular 
importance in the field of modern electronics.  
Ferroelectric materials that show spontaneous electric dipoles 
below a transition temperature $T_c$ have been used 
in applications such as high-density, non-volatile memories.\cite{review2}
In conventional ferroelectric materials, 
two kinds of electric dipoles are recognized. 
One is a displacement type, in which cations shift relative 
to anions below $T_c$. 
The other is an order-disorder type, 
in which polar molecules are randomly oriented above $T_c$, 
and align below $T_c$. BaTiO$_3$ and NaNO$_2$ are typical examples 
of the former and the latter types, respectively.\cite{review1}
Recently, a third type of ferroelectricity was noted 
in the layered iron oxide LuFe$_2$O$_4$, 
in which the electric dipole comes 
from the ordering of the  Fe$^{2+}$ and Fe$^{3+}$ 
ions on a double-layered triangular lattice.\cite{ikeda2005}
These dipoles are understood in terms of atomic positions 
in the crystal of interest.

Here we show a truly electronic type 
of electric dipole in the organic salt 
$\kappa$-(BEDT-TTF)$_2$Cu$_2$(CN)$_3$, 
where BEDT-TTF stands for bis(ethylenedithio)-tetrathiafulvalene.
$\kappa$-(BEDT-TTF)$_2$Cu$_2$(CN)$_3$ is known 
as a dimer Mott insulator,\cite{geiser1991} 
and has been  investigated as an ideal candidate 
for a spin-liquid state.\cite{shimizu2003,yamashita2008,yamashita2009} 
This organic salt is a layered  compound 
in which the BEDT-TTF and Cu$_2$(CN)$_3$ layers 
are alternately stacked along the $a$ axis. 
The BEDT-TTF layer is responsible for the electrical 
and magnetic response, 
while the Cu$_2$(CN)$_3$ layer only acts 
to electro-statically stabilize the crystal. 
The Greek letter $\kappa$ specifies a packing pattern 
of the BEDT-TTF molecules, which is schematically shown in Fig. 1(a). 
In this pattern, hole exists per two dimerized molecules 
as indicated by the red dotted ellipsoids. 
Thus, if one regards the two molecules as a sort of ``atom'' 
represented by the closed circles in Fig. 1(b), 
one can identify this BEDT-TTF layer with a ``half-filled'' system 
in which one hole is localized on each site.\cite{kino1995} 
This system meets the definition of a Mott insulator, 
hence the name ``dimer Mott insulator.'' 
The spin degrees of freedom on each localized hole acts 
as a magnetic dipole as shown in Fig. 1(c). 
The magnetic dipoles interact via a superexchange interaction 
$J/k_B \sim$250 K, but an antiferromagnetic transition 
does not take place above 32 mK, 
owing to the frustration coming from the geometry 
of the triangular lattice based on dimers. 
%\color{red}
To be more precise, recent ab-initio calculations reveal that 
the dimer network cannot be
regarded as a regular triangle in which the anisotropy 
of the transfer integral reaches 0.8, but still
the spin system is in the frustrated region.\cite{kandapal,nakamura}
%\color{black}
This lack of long range order of the magnetic dipoles 
defines a spin liquid.\cite{shimizu2003}

We have found that $\kappa-$(BEDT-TTF)$_2$Cu$_2$(CN)$_3$
exhibits dielectric relaxation below around 60 K,
which is not expected from the charge excitations in
conventional Mott insulators.
Based on the extended Hubbard model, we ascribe this
to the charge disproportionation 
within the dimer driven by 
the intersite Coulomb interaction.

\begin{figure}[t]
  \begin{center}
   \includegraphics[width=8cm,clip]{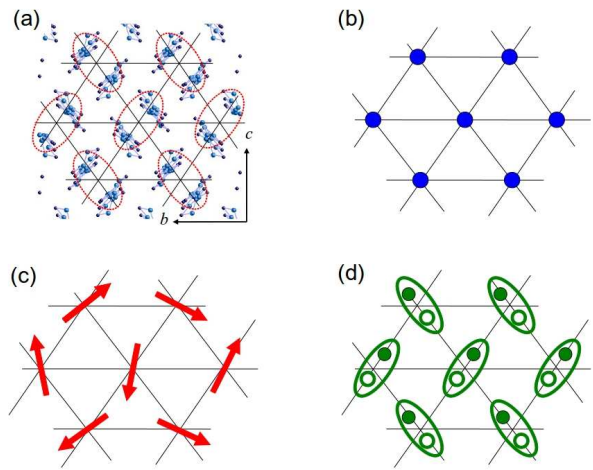}
   \caption{
   (Color online)
   (a) Schematic of the BEDT-TTF layer in 
   $\kappa$-(BEDT-TTF)$_2$Cu$_2$(CN)$_3$. 
   The dotted ellipsoids represent dimerized molecules. 
   (b) A triangular lattice of dimers, 
   where closed circles represent identified with the
   dimerized molecules. 
   (c) A triangular lattice of magnetic dipoles (spins). 
   (d) A triangular lattice of electric dipoles, where open and
   closed circles represent positive and negative point charges,
   respectively. 
   This is identical to Fig. 5(f).
   }
   \label{fig1}
  \end{center}
\end{figure}

\section{Experiemental}
$\kappa$-(BEDT-TTF)$_2$Cu$_2$(CN)$_3$ was grown 
by an electrochemical method. 
Dielectric constant ($\varepsilon$) 
and resistivity ($\rho$) measurements were 
carried out using an HP4284A impedance analyzer 
along with cooling down in a liquid-helium cryostat. 
Electric displacement-electric field ($D-E$) 
curves were measured with a homemade apparatus based 
on the Sawyer-Tower circuit \cite{st} 
with a maximum electrical field of 1000 V.
The measurement direction was set to be 
perpendicular to the BEDT-TTF layer (along the $a$ axis),
because a dielectric constant cannot be measured
precisely for conductive media.\cite{nagao}
The cross-plane resistivity is much higher than
the in-plane one, and the contact resistance was safely neglected. 
%\color{red}
We should note that the temperature and frequency dependence of $\varepsilon$
is similar between the in- and cross-plane directions
of layered materials,
so that the ac response of a two-dimensional dielectric material
can be discussed from the cross-plane measurement 
at least qualitatively.\cite{takayanagi2002}
The magnitudes of $\varepsilon$ are even close in some materials
such as LuFe$_2$O$_4$.\cite{ikeda2000}
%\color{black}

\begin{figure}[t]
  \begin{center}
   \includegraphics[width=8cm,clip]{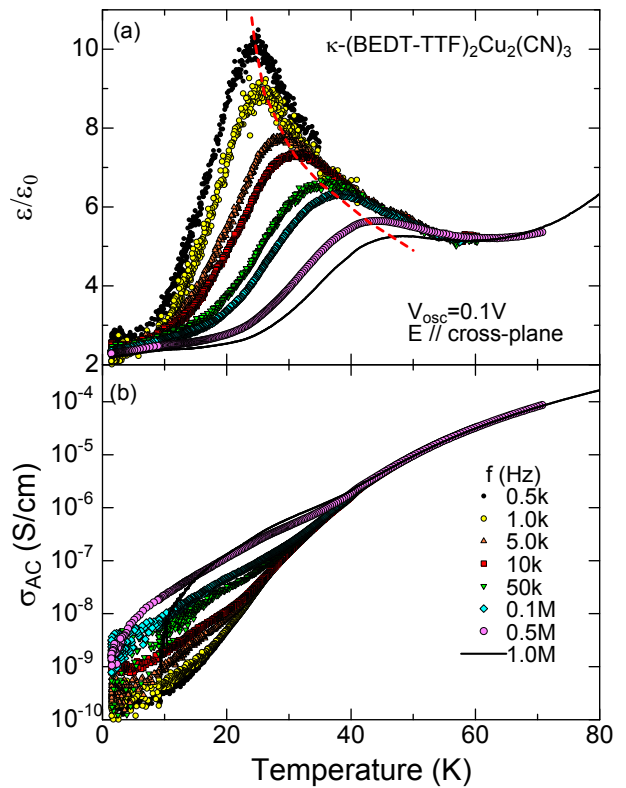}
   \caption{
   (Color online)
   (a) The dielectric constant of a single crystal of 
   $\kappa$-(BEDT-TTF)$_2$Cu$_2$(CN)$_3$
   along the a-axis (cross-plane) direction 
   at various frequencies as a function of temperature. 
   The randomly oriented electric dipoles appear below 60 K. 
   The dotted curve indicates the peak temperature $T_{\rm max}$.
   (b) The AC electrical conductivity of the same
   crystal. 
   }
   \label{fig2}
  \end{center}
\end{figure}

\section{Results and Discussion}
Figure 2(a) shows the dielectric constant with respect to 
various frequencies plotted as a function of temperature. 
The dielectric constant increases with decreasing 
temperature below 60 K, and simultaneously begins 
to show frequency dependence.  
As temperature is lowered, 
the dielectric constant goes through a broad maximum 
at a temperature $T_{\rm max}$ depending 
on the measurement frequency $f$, 
then decreases toward 2.1-2.5. 
$T_{\rm max}$ corresponds to a crossover temperature 
below which the response to the changing electric field 
begins to lag. 
The  AC conductivity also shows 
frequency dependence as shown in Fig. 2(b). 
%\color{red}
In this case, the frequency dependence becomes significant
below around 40 K, which is lower than that for the dielectric constant.
Roughly speaking, we see that the ac conductivity bends around
$T_{\rm max}$ and remains higher than the conductivities at lower frequencies.
%\color{black} 

The dielectric relaxation we observe is indeed unconventional. 
First, the charge degrees of freedom is believed 
to be insubstantial in the Mott insulator, 
but the increasing dielectric constant 
below 60 K indicates the existence of randomly oriented 
electric dipoles as shown in Fig. 1(d). 
Secondly, the dielectric relaxation implies collective motion 
of the electric dipoles. 
If all the dipoles were independent, 
the response would be independent of $f$, because $k_BT \gg hf$.

\begin{figure}[t]
  \begin{center}
   \includegraphics[width=6cm,clip]{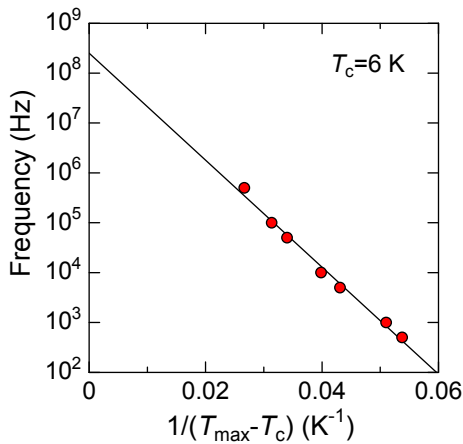}
   \caption{
   (Color online)
   The measurement frequency $f$ 
   plotted as a function of $1/(T_{\rm max}-T_c)$.
   $T_{\rm max}$ is the peak temperature
   at which the dielectric constant goes
   through a broad maximum, and
   $T_c$ is assumed to be 6 K.
   The solid line corresponds to the fitting curve,
   where $f_0$ and $E_g/k_B$ are evaluated to be
   $2.5\times 10^8$ Hz and 250 K, respectively.
   }
   \label{fig3}
  \end{center}
\end{figure}

%\color{red}
Similar relaxation behavior is widely
observed in disordered systems such as glass-forming liquids,
spin-/cluster-glasses,
and relaxor ferroelectrics.
The measurement frequency is plotted as a function
of $1/(T_{\rm max}-T_c)$ in Fig. 3.
$T_c$ is the transition temperature, which
we assume to be 6 K because the 
specific heat,\cite{yamashita2008} 
the thermal conductivity,\cite{yamashita2009}
and the thermal expansion coefficient\cite{manna2010}
show anomaly at this temperature.
The data are roughly linear, which suggests that
the Vogel-Fulcher law $f =f_0 \exp[-E_0/k_B(T_{\rm max}-T_c)]$ 
consistently explains the frequency dependence of $T_{\rm max}$. 
Since the Vogel-Fulcher law is widely observed in
glass-forming liquid,\cite{glass}
spin-/cluster-glasses,\cite{spinglass,clusterglass} 
and relaxor ferroelectrics,\cite{bakov2006,viehland}
it is natural to conclude that the dielectric relaxation 
comes from disordered arrangements of the electric dipoles.

The values of $f_0$ and 
$E_0/k_B$ are evaluated to be 
$2.5\times 10^8$ Hz and 250 K, respectively.
We should emphasize that $f_0$ is significantly 
smaller than a typical value of $f_0\sim 10^{12}$ Hz 
for relaxor ferroelectric materials.\cite{viehland}
The physical meaning of $1/f_0$ is a typical time scale 
for ac response in the high temperature limit,
which should be longer in domain motions than in individual motions.
It is known for magnetic systems that cluster glass materials 
(where ferromagnetic domains are randomly oriented) 
tend to show smaller values of $f_0$ than 
spin glass materials (where individual spins 
are randomly oriented).\cite{clusterglass}
In this respect, we think that the polar domains
are disordered rather than the individual electric dipoles.
Similar disorder is reported in magnetic resonance 
experiments,\cite{shimizu2006,kawamoto2006} and may be related 
to the suppression of the long range order of magnetic moments. 

It is well known that disorder in the terminal ethylene group
seriously affects the physical properties 
in some (BEDT-TTF)-type organic salts.
Some physical quantities such as superconducting $T_c$ significantly 
depend on the cooling rate.\cite{kanoda}
Theoretically, this type of disorder can be
a pair breaker of superconductivity,\cite{powell2004} 
and modifies the intra-dimer Coulomb interaction.\cite{powell2009}
We performed dielectric measurements with different
cooling rates ranging from 0.5 to 10 K/min,
and found that the data were essentially the same 
as in Fig. 2 (not shown).
We further measured the dielectric response for
a deuterated sample, and found that the data
were again essentially the same (not shown).
These results indicate that the observed dielectric
relaxation does not come from the disordered arrangements
of the hydrogen bonding in the terminal ethylene group.
In this context, we can say that the single crystal used here
contains no substantial lattice defects or disorder, 
as pure as crystals of other organic salts. 
In relaxor ferroelectric materials, in contrast,
more than a few percents of the host atoms 
are replaced by different atoms, 
which can be seeds for the inhomogeneity.\cite{bakov2006,horiuchi2000}  

\begin{figure}[t]
  \begin{center}
   \includegraphics[width=8cm,clip]{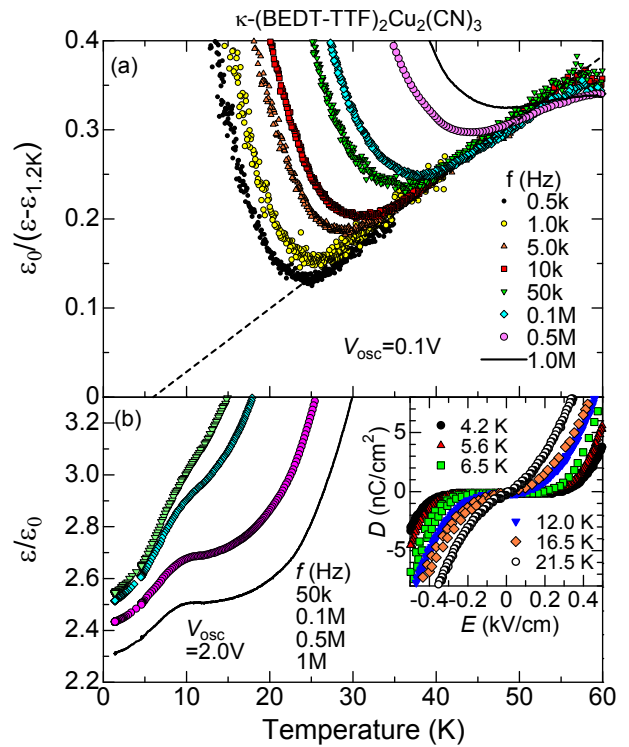}
   \caption{
   (Color online)
   (a) The inverse dielectric constant plotted as a function of
   temperature. 
   The temperature-independent part 
   (the dielectric constant at 1.2 K) has been subtracted. 
   The dotted line indicates that the dielectric constant 
   obeys the Curie-Weiss law with a Curie temperature 
   $T_c=$6 K.
   (b) The low-temperature part of the dielectric constant. 
   A frequency-independent cusp is observed near $T_c$. 
   The inset shows the electric displacement $D$ 
   plotted as a function of external electric field $E$.
   }
   \label{fig4}
  \end{center}
\end{figure}

%\color{black}
Although no clear phase transition is observed 
in the thermodynamic quantities 
of this material,
one may find a trace of a transition temperature.
Figure 4(a) shows the inverse dielectric constant 
as a function of temperature. 
These data are plotted after subtracting 
the dielectric constant at 1.2 K as the temperature-independent part. 
The dielectric constant below 60 K obeys the Curie-Weiss law, 
i.e., it is roughly inversely proportional to $T-T_c$ 
with 
$T_c=$6 K.
As shown in Fig. 4(b), around $T_c$, the dielectric constant 
has an anomaly that is almost independent of frequency. 
As shown in the inset of Fig. 3(b), 
the electric displacement $D$ shows 
no remnant polarization below 6 K, 
indicating that the ordering of the electric dipoles 
is of antiferroelectric type. 

\begin{figure}[t]
  \begin{center}
   \includegraphics[width=8cm,clip]{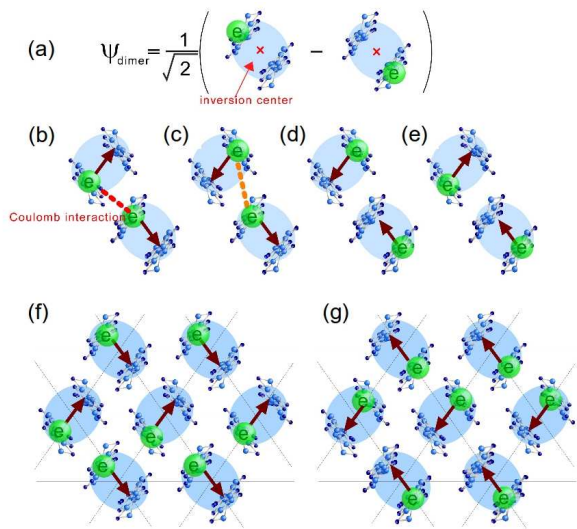}
   \caption{
   (Color online)
   (a) Schematic of a quantum electric dipole. 
   (b)-(e) Schematics of two neighboring quantum 
   electric dipoles. 
   The arrows represent the electric dipoles. 
   (b) and (c)  represent the cases when two electrons 
   are come close to one another, whereas (d) and (e) represent the
   cases when the two electrons are far apart. 
   (f) and (g) Possible short-range domains 
   of the electric dipoles fluctuating collectively.
   }
   \label{fig5}
  \end{center}
\end{figure}

We propose a microscopic origin of this antiferroelectric response. 
The extended Hubbard model based on the BEDT-TTF molecular 
orbitals is a widely accepted model 
for this family of organic materials, 
which consists of transfer integral, on-site Coulomb repulsion, 
and inter-site Coulomb repulsion terms.\cite{seo2004}  
Owing to the large transfer integrals 
and Coulomb repulsion between the dimerized molecules, 
an electron on a dimer is described by the superposition 
of two quantum states expressed by 
$\psi_{\rm dimer} = a_A\phi_A +a_B\phi_B$
with  the coefficients $a_A =a_B = 1/\sqrt{2}$  
as shown in Fig. 4(a).  
Here, $\phi_A$ and $\phi_B$  represent the wavefunctions 
of the dimerized A and B molecules. 
The electron on the dimer thus fluctuates between A and B, 
rather than staying statically at the center of the dimer.
On the other hand, the inter-dimer Coulomb repulsion 
polarizes the dimers (i.e. unequalizes $a_A$ and $a_B$)
such that the electrons on neighboring dimers stay apart 
as far as possible. 
When these two electrons come close [Figs. 4(b) and (c)], 
the repulsion is higher than when they are far apart 
[Figs. 4(d) and (e)]. 
In this way, the intra-dimer quantum fluctuation 
and the dipole-dipole interaction compete.
%\color{red}
In related organic conductors of one dimension,
the same sort of interaction induces a ferroelectric transition.\cite{nad2006}
%\color{black}
Owing to the zig-zag packing of BEDT-TTF molecules, 
the dipole-dipole interaction becomes relatively small, 
and concomitantly the intra-dimer quantum fluctuation is dominant, 
allowing the effects of dipole-dipole interaction 
to remain as short range correlations. 
As a result, the system goes back and forth between two configurations;
Figs. 4(f) and (g) show two snapshots of the quantum fluctuation coupled
with the inversion symmetry. 

%\color{red}
Our model suggests that the charge ordering instability 
survives in $\kappa$(BEDT-TTF)$_2$Cu$_2$(CN)$_3$. 
This is reasonable; depending on the degree of dimerization, 
the ground state of the quarter-filled
organic salts can be continuously changed from 
the charge ordered insulator to the dimer Mott insulator.\cite{seo2004,seo2006} 
These two pictures are extreme limits, and real materials 
lie in between. 
Detailed theoretical study on this picture is written separately,\cite{hotta} 
based on the model including the transverse 
Ising term which accounts for the charge degrees of freedom, 
and the Kugel-Khomskii-term describing the couplings of spin and charges. 
Motivated by our experiment, Naka and Ishihara,\cite{ishihara2010} 
in parallel with Ref. \onlinecite{hotta}, 
also calculated the mean-field phase diagram on the similar model, 
and successfully explained a possible existence of a ferroelectric charge order.
It should be noted that before our experimental study, 
Clay and his coworkers \cite{clay2010} have discussed 
the physical properties of the organic salts in terms of 
electron-paired crystal, and already predicted that the charge ordering 
pattern is hidden in the title compound.
Actually, their predicted pattern is similar to Figs. 5(f) and 5(g). 
%\color{black}

We can understand the dielectric response qualitatively 
using the above concept.  
This electric dipole is tightly bound to the molecular arrangement, 
fluctuating collectively within a certain length scale. 
When an external electric field is applied, such collective
dynamical domains easily obey the external field,  retaining an
inhomogeneous nature. 
Thus a mean square value of $\sqrt{<q^2>}L$ is induced
by the external field, 
where $q$ is proportional to the difference of the
electron densities $n_A=|a_A|^2$ and $n_B=|a_B|^2$ 
and $L$ is the distance between the A and B molecules. 
From the slope in Fig. 3(a), 
$\sqrt{<q^2>}$ is evaluated to be 0.1$e$. 
%\color{red}
Of course, the above estimate of $L$ and $\sqrt{<q^2>}$ 
came from the oversimplified picture. 
In the present experiment, the dipole moments tend to align
perpendicular to the BEDT-TTF layer, and the charge should be
polarized along the BEDT-TTF molecule.
We should note that the the charge distribution of 
the $\pi$ electrons on the molecule is about $5-10$ \AA,
which is the same length scale as the inter-molecular distance.
Thus we believe that the estimated charge 
disproportionation of 0.1$e$ will not be off the mark.
%\color{black}

Finally, we briefly add some notes on 
the nature of this dielectric relaxation.
(i) The dielectric constant at 1 MHz above 10 K is independent
of DC bias up to 2 kV/cm.
This makes a remarkable contrast to 
relaxor ferroelectric materials \cite{kutajak2006} 
or internal barrier-layer capacitors,\cite{ccto}
and excludes a possibility for extrinsic origins \cite{wang}  
(ii) The dielectric response is also 
independent of magnetic field up to 15 T, 
and represents a remarkable contrast to multiglass or multiferroic
materials.\cite{multiferro,multiglass} 
(iii) We observed similar dielectric relaxation 
in $\beta^\prime$-(BEDT-TTF)$_2$ICl$_2$.\cite{kobayashi1986}
This clearly indicates that this type of electric dipoles 
widely exists in dimerized BEDT-TTF molecules, 
and  reveals the importance of the charge degrees 
of freedom in such systems.

\section{Summary}
In summary, we have measured the dielectric constant of
the dimer Mott insulator $\kappa-$(BEDT-TTF)$_2$Cu$_2$(CN)$_3$,
and have found anomalous dielectric relaxation below 
around 60 K.
This relaxation resembles that of relaxor ferroelectric materials, 
which strongly suggests the existence
of interacting electric dipoles.
This electric dipole is ascribed to the charge disproportionation
within the BEDT-TTF dimers driven by the intermolecular
Coulomb interaction.
We have evaluated the charge disproportionation to be 0.1$e$
from the temperature dependence of the dielectric constant. 
The present results indicate that nontrivial charge degrees 
of freedom survive in the dimer Mott insulator,
which may be related to the characteristic properties
of this family such as pressure-induced superconductivity.

The authors would like to thank Y. Nogami, Y. Noda, 
S. Ishihara, M. Lang and M. de Souza for fruitful discussion.
This work is partially supported by the Grant-in-Aid 
for Scientific Research 
(No. 21340106, 20340085 and 21110504) and 
JSPS Research Fellowships.

\end{document}